\documentclass[a4paper,fleqn]{cas-sc}

\usepackage[numbers]{natbib}

\usepackage{xspace} 
\newcommand{\pryso}[0]{{Pr$^{3+}$:Y$_2$SiO$_5$}\xspace}
\newcommand{\eryso}[0]{{Er$^{3+}$:Y$_2$SiO$_5$}\xspace}

\newcommand{\yso}[0]{Y$_2$SiO$_5$\xspace}
\newcommand{\er}{Er$^{3+}$\xspace}

\newcommand{\diff}[1]{{#1}}

\begin{document}
\shorttitle{Site-selective angular-resolved absorption spectroscopy of erbium}
\shortauthors{Y. Petit et~al.}

\title[mode = title]{Demonstration of site-selective angular-resolved absorption spectroscopy of the $^{4}I_{15/2} \rightarrow $$^{4}I_{13/2}$ erbium transition in the monoclinic crystal Y$_2$SiO$_5$ }

\author[1,2]{Yannick Petit}
\author[3]{Beno\^it Boulanger}
\author[3]{J\'er\^ome Debray}
\author[3]{Thierry Chaneli\`ere}[orcid=0000-0003-1715-9520]

\address[1]{Universit\'e de Bordeaux, CNRS, ICMCB, UMR 5026, F-33608 Pessac, France}
\address[2]{Universit\'e de Bordeaux, CNRS, CEA, CELIA, UMR 5107, F-33405 Talence, France}
\address[3]{Univ. Grenoble Alpes, CNRS, Grenoble INP, Institut N\'eel, 38000 Grenoble, France}





\begin{abstract}[S U M M A R Y]
We study the angular dependence in polarized light of the optical absorption for the \er transition $^{4}I_{15/2} \rightarrow $$^{4}I_{13/2}$ in \yso, revealing thus the associated anisotropy and the orientation of the related absorption principal directions in the dielectric plan perpendicular to the monoclinic axis b. The measurements are performed at low temperature. This allows us to isolate the lowest crystal field levels in the ground and excited states. We spectrally resolve and independently characterize the two yttrium substitution sites in the \yso matrix. The absorption tensor components cannot be unambiguously determined yet while only considering the investigated dielectric plane. Still, measurements remarkably demonstrate that this transition of interest well resolved at low temperature is not only a magnetic-dipole allowed transition but indeed a hybrid electric-magnetic transition.
\end{abstract}




\maketitle

\section{Introduction}
The interest for high resolution spectroscopy in rare-earth samples have been recently renewed by the active quest for quantum photonics devices \cite{Goldner_review}. Lower doping levels are generally targeted, 50 ppm in our case, as compared to those for laser applications but the host matrices are the same, benefiting from years of development to improve crystal quality. Among them, the monoclinic yttrium orthosilicate (\yso) is certainly the most studied, with an unprecedented record quantum coherence time for the europium dopant \cite{zhong2015optically}. The low symmetry of both the crystal matrix and involved substitutional sites lead to a strong anisotropy being then exploited to tune the spin properties by accurately adjusting the magnetic field in the crystal frame \cite{bottger_effects_2009, zhong2015optically, ranvcic2018coherence}. Erbium shows remarkable performances in \yso in terms of coherence \cite{bottger_effects_2009, ranvcic2018coherence}. Among the other lanthanides, \er holds a lot of promises because of its compatibility with the fiber telecom range which opens many perspectives in classical and quantum processing. Major effort has thus been made to fully understand the orientation of the magnetic properties. Using advanced spectroscopic studies, the anisotropic tensors describing the spin are accurately known, including the crystal-field \cite{li1992spectroscopic, PhysRevLett.123.057401}, the different g-tensors \cite{sun_magnetic_2008} (in both substitution sites of yttrium and in the ground and optically excited state of \er). The hyperfine tensors for odd-isotopes are also known \cite{PhysRevB.74.214409, PhysRevB.97.024419}.

Even if the optical $^{4}I_{15/2} \rightarrow $$^{4}I_{13/2}$  transition is put forward to stimulate the applications, the orientation of the absorption principal directions as well as the associated principal values are essentially unknown. Absorption is indeed a key parameter for the efficiency of the different optical functions currently under investigation, like quantum memories \cite[see][and references therein]{Heshami, CHANELIERE201877} or the more recent opto-RF quantum transduction \cite[see][and references therein]{Lambert2020, Lauk_2020}. Despite its importance, the absorption is only known along the so-called extinction axes, which corresponds to the principal axes of the  dielectric tensor \cite{liu2006spectroscopic}. However, the off-diagonal components of the absorption tensor are not known. The impact on the propagation of the mismatch between the absorption and dielectric tensors has been already addressed in \pryso. This pioneering work also performed at low temperature, where all the absorption lines are resolved, corresponds to a simplified situation because the absorption lies in a principal plane of the dielectric tensor \cite{kinos,kinos_phd}. \diff{In any case, a complete characterization requires to explore the different crystalline directions, away from the principal axes, in order to access the maximal absorption by properly cutting and orientating the promising  \yso crystal.}

A complete determination of the optical properties in \mbox{monoclinic} crystals is not obvious and requires an accurate control of the sample orientation and incoming polarization. Despite implications in laser physics and non-linear optics, and even though the fundamental background is well established for electric-dipolar transitions, experimental data are quite rare \cite{kaminskii1979anisotropy, owen1998orientation, petit2013recent}. 

The spherical crystal shape is ideal to explore all possible orientations with a constant propagation length \cite{petit2013recent}. However, a full three dimensional rotation of a sphere is not compatible with measurements in our cryostat and it appears technically challenging at low temperature. As the cylindrical geometry (with its revolution axis along the crystallographic {\bf b}-axis) may be sufficient for monoclinic crystals \cite{dinndorf1992principal}, we choose this configuration for our experiment performed at cryogenics temperature. Still, as further detailed in section \ref{sec:analysis}, the cylindrical geometry won't be sufficient to fully reconstruct the absorption tensor of the  $^{4}I_{15/2} \rightarrow $$^{4}I_{13/2}$ transition. Nevertheless, the low temperature measurements give an unprecedented spectral resolution to characterize unambiguously the different crystal field levels of the two substitutional sites as detailed in section \ref{sec:setup}. We will first present the experiment and then discuss the characterization of the absorption tensor. As we will see, we cannot fully distinguish between the electric-dipole (so called forced ED) and magnetic dipole (MD) contributions, but we can only bound their relative weight. The MD transition is indeed allowed for the the $^{4}I_{15/2} \rightarrow $$^{4}I_{13/2}$ of \er \cite{li1992spectroscopic}.

\section{Optical setup}\label{sec:setup}

\begin{figure}[htbp]
\centering
\includegraphics[width=.6\textwidth]{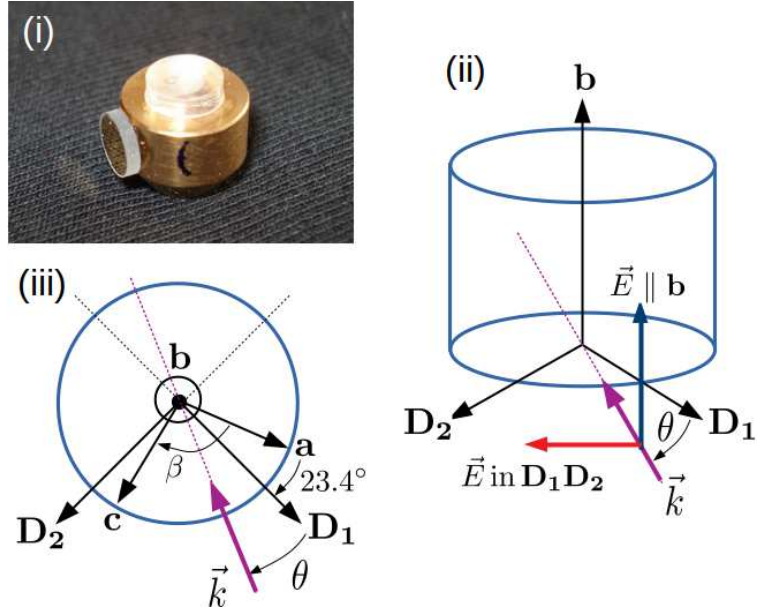}
\caption{(i): a polished cylinder is centered on a brass holder and placed on a rotating mount in the cryostat. A small mirror is glued on the side of the brass holder, serving as a common orientation reference (laser auto-collimation) between the Laue diffractometer and the measurement setup. (ii): reference dielectric frame ({\bf b,D$_2$,D$_1$}). The angle $\theta$ is defined as the rotation angle from {\bf D$_1$} toward {\bf D$_2$}. (iii) Projection and identification of the axes in the monoclinic plane perpendicular to {\bf b}. In practice, the propagation direction $\vec{k}$ is kept constant while rotating the sample around {\bf b}. The electric field polarization can be either aligned with {\bf b} (blue) or in the ({\bf D$_1$,D$_2$}) plane (red).}
\label{fig:schema}
\end{figure}

Following the method in \cite{Menaert:17}, we prepare a polished cylinder of \yso (thickness 3 mm, diameter 5.8 mm) doped with an \er atomic concentration of 50 ppm grown by Scientific Materials \diff{(Fig.\ref{fig:schema}, (i))}. The {\bf b}-axis, the cylinder axis, is fixed by monochromatic diffraction ($\pm0.1^\circ$ precision) in order to properly shape the cylinder in the ({\bf a,c}) plane (cylindricity error under 1\%). Finally, the exact in-plane orientation is determined with a Laue diffractometer with a sub degree precision.
We chose this configuration because the {\bf b}-axis, also called the special monoclinic axis, defines the axis of the dielectric frame that remains fixed as a function of any dispersive parameter of the dielectric permittivity, like the wavelength or the temperature for example. The two other axes, perpendicular to the {\bf b}-axis, will be called {\bf D$_1$} and {\bf D$_2$} in the following, and may vary as a function of wavelength and temperature. The other reason is that the special axis is a principal axis for both the real and imaginary part of dielectric permittivity \cite{petit2013recent}.

The cylinder is placed in a variable-temperature-insert liquid helium cryostat.  The rotation is controlled manually from room-temperature though an O-ring sealed rotating pole. The precision is 2$^\circ$, which corresponds to the rotating mount scale increment.
%
We collimate the beam in the cylinder by properly focusing the incoming beam using cylindrical lenses with a  focal length of 75 mm at the input and output of the cryostat. We place a polarizing cube and half-wave plate to control the incoming polarization to independently address each of the two polarization Eigenmodes, i.e. parallel or orthogonal to the {\bf b}-axis.

\begin{figure}[htbp]
\centering
\includegraphics[width=.6\linewidth]{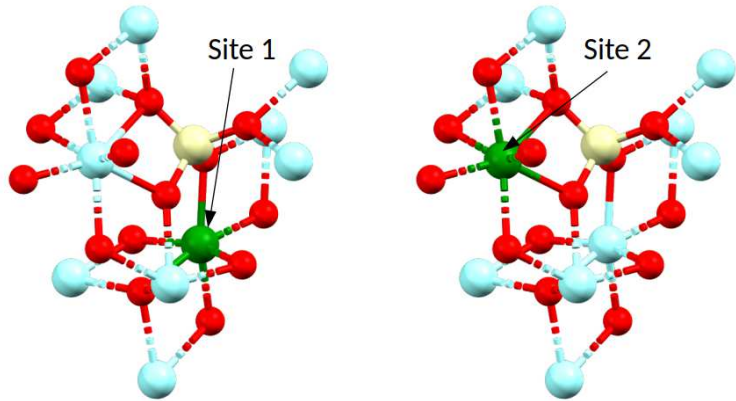}
\caption{Reduced crystal cell representing the two possible substitutional sites in the \yso matrix: site 1 (left) and site 2 (right) with 6-coordinate and 7-coordinate respectively. Colour code: cyan - yttrium, red - oxygen, yellow - silicon and green - erbium substituting an yttrium.}
\label{fig:sites}
\end{figure}

Our objective is to scan the $^{4}I_{15/2} \rightarrow $$^{4}I_{13/2}$ transition between the lowest crystal field levels of both the sites 1 (6-coordinate) and 2 (7-coordinate) which represent the two possible substitutional sites in the \yso matrix (see Fig.\ref{fig:sites}). \diff{The exact ratio between both sites cannot be determined quantitatively because a slightly weaker substitution could be potentially compensated by a larger oscillator strength for example. That being said, \er is abundantly present in both sites having the same valency and size. The crystal fields are very similar which is confirmed by very similar optical absorption values \cite{KURKIN1980233}. So the sites occupancy seems to be well balanced, especially at low doping level.}

\begin{figure}[htbp]
\centering
\includegraphics[width=.6\textwidth]{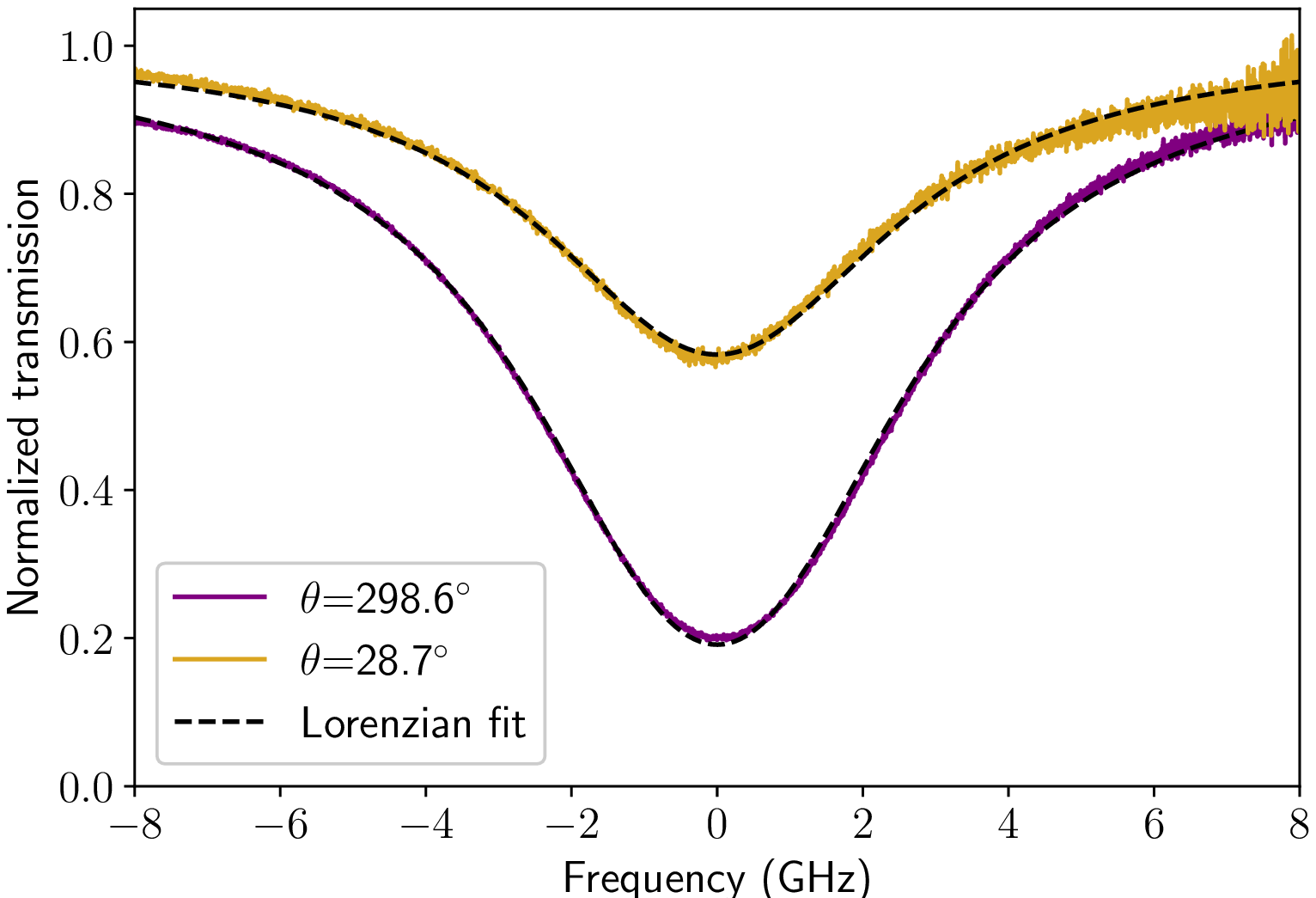}
\caption{\diff{Examples of the transmission spectrum for the  $^{4}I_{15/2} \rightarrow $$^{4}I_{13/2}$ transition of site 1 (1536.5 nm). Theses spectra correspond to the maximum and minimum optical depths of $1.65$ and $0.54$ respectively for an angular rotation $\theta=298.6^\circ$ (purple) and $\theta=28.7^\circ$ (orange) respectively when the polarization is aligned with {\bf b}. The dashed lines are Lorentzian fits of the experimental curves.}}
\label{fig:Two_spectra_article}
\end{figure}

Then we use the same fibred DFB laser (slow diode Peltier device temperature tuning) centered at 1536.5 nm and 1538.9 nm, for both sites respectively. We operate the cryostat at 45 K. At this temperature, the measured optical depth corresponding to the natural logarithm of the transmission is of the order of one, the transition linewidths being $\sim$ 2 GHz. At lower temperature, the linewidth is lower ($\sim$ 0.5 GHz) thus increasing the absorption to a point where the transmission is becoming too weak to be accurately measured.
We scan the laser frequency across the resonance to record a complete line spectrum for each absorption measurement \diff{as illustrated in Fig.\ref{fig:Two_spectra_article}}. The natural logarithm of the transmission spectrum is fitted by a \diff{Lorentzian} curve whose height gives the optical depth.

\section{Experimental results}

As reference frame for the measurement, we use the dielectric frame ({\bf b,D$_2$,D$_1$}). This notation has been introduced for the first spectroscopic measurements of \eryso \cite{li1992spectroscopic} and is widely used to reference the spin Hamiltonian components. The  ({\bf b,D$_2$,D$_1$}) frame is the dielectric frame ({\bf X,Y,Z}), these axes respectively bearing the low, intermediate and large principal indices of refraction \cite[p.386]{liu2006spectroscopic, Woodburn, sellmeier}. The relative orientation of  ({\bf b,D$_2$,D$_1$}) and  ({\bf a,b,c}) is summarized in Fig.\ref{fig:schema}. The dielectric frame slightly rotates around the {\bf b}-axis as a function of wavelength, as observed in \cite{Traum:14}. Moreover, such a rotation of the dielectric frame orientation becomes negligible in the infrared \diff{(less than 1$^\circ$)}, allowing to consider that this orientation offset is within our measurement uncertainty (precision 2$^\circ$). Therefore, we did not try to remeasure the exact positions {\bf D$_2$} and {\bf D$_1$} at our wavelength and we simply fix the position of  {\bf D$_1$}  at $23.4^\circ$ from {\bf a}, as chosen \diff{in Fig.(5.a) of \cite{Traum:14}}. \diff{This recent reference value is in good agreement with the previous measurement of a $23.8^\circ$ offset for {\bf D$_1$} \cite{li1992spectroscopic}. In any case, the slight variations of the optical frame orientation are smaller than our measurement precision. Fixing the {\bf D$_2$} and {\bf D$_1$} with respect to the crystal frame allows the crystal to be oriented using X-ray diffraction without depending on the wavelengths or dopants of interest and to keep a common reference for the g-tensor definitions \cite{sun_magnetic_2008}}. The monoclinic angle between {\bf a} and {\bf c} is $\beta=102.65^\circ$. 

\begin{figure}[htbp]
\centering
\includegraphics[width=.8\textwidth]{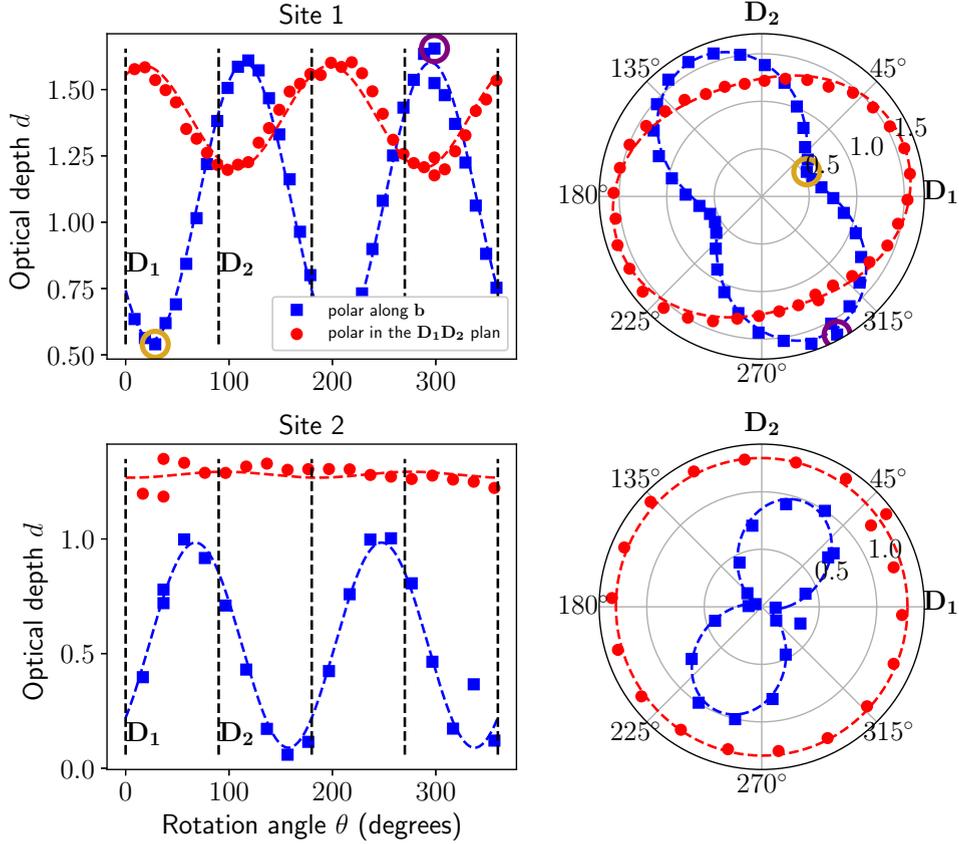}
\caption{Angular distributions of optical depths for site 1 (top) and 2 (bottom) versus the cylinder orientation $\theta$, for the two Eigenmode polarizations either aligned with {\bf b} (blue) or in the ({\bf D$_1$,D$_2$}) plane (red). When $\theta=0$, then $\vec{k} \parallel $ {\bf D$_1$}.  When $\theta=90^\circ$, then $\vec{k} \parallel $ {\bf D$_2$}.
Right panel: same data in polar representation. The blue and red dashed curves are fitted with \eqref{eq:dinndorf}. \diff{The two measurements represented in Fig.\ref{fig:Two_spectra_article} have been marked with purple ($\theta=298.6^\circ$) and orange ($\theta=28.7^\circ$)  circles for site 1.}} 
\label{fig:ErYSO_all_article}
\end{figure}

As we rotate the sample by an angle $\theta$ from {\bf D$_1$} to {\bf D$_2$}, the polarisation is either constant along ${\bf b}$ or it lies in ({\bf D$_1$,D$_2$}) plane, respectively corresponding to the ordinary or extraordinary polarization eigenmodes, as defined in Fig.\ref{fig:schema}. In the latter case, the polarization rotates in the plane as well (along a direction close to $\theta$+90$^\circ$, while neglecting the minor influence of double refraction). The measurement of the optical depth is reproduced for site 1 and site 2, and results are plot in Fig.\ref{fig:ErYSO_all_article}.

The different angular distributions in polarized light exhibit contrasted modulations, expect for site 2 when polarization lies in the ({\bf D$_1$,D$_2$}) plane. As a signature of the local anisotropy in the \yso matrix, well-known for the different components of the spin Hamiltonian, the absorption varies a lot (see for site 2, polarisation along {\bf b} in Fig.\ref{fig:ErYSO_all_article}). \diff{As an illustration, the g-factor varies by a factor of 10 to 20 by comparing the maximum and minimum values at different magnetic field orientations for site 2 \cite{sun_magnetic_2008}. We retrieve a comparable anisotropy ratio for the absorption of the same site (blue curve in Fig.\ref{fig:ErYSO_all_article}, site 2).}

The absorption eigenaxes (maximum and minimum) that we exhibit by rotating the cylinder around {\bf b} do not coincidence with the  dielectric frame ({\bf b,D$_2$,D$_1$}) axes (see polar plots in Fig.\ref{fig:ErYSO_all_article}) as expected \cite{petit2013recent, dinndorf1992principal}. Moreover, it is remarkable to observe that the absorption eigenaxes from site 1 and site 2 are different. This results indeed from the different local environments of the two sites, and primarily the different coordination numbers, with oxygen in that case. Beyond the qualitative analysis of the angular distribution, the data can be discussed further by noting first that \yso belong to the monoclinic class where {\bf b} is a fixed and common eigendirection to any optical tensors by symmetry, allowing the dielectric frame as well as any resonant absorbing or emitting frame to show orientation dispersion around {\bf b}  \cite{petit2013recent}; and secondly that the $^{4}I_{15/2} \rightarrow $$^{4}I_{13/2}$ is MD allowed.

\section{Analysis}\label{sec:analysis}
%

The theoretical analysis of absorption anisotropy relies on symmetry considerations \cite{aminov1985theory, Stedman, petit2013recent}. The key point for a monoclinic crystal such as \yso is the fact that {\bf b} is an axis of symmetry.

To start up, as judiciously noted in \cite{dinndorf1992principal}, the transition dipole follows the general projection rules of a spin-1/2 particle submitted to an anisotropic Hamiltonian. In short, the absorption angular distribution can be decomposed as 

\begin{align}\label{eq:dinndorf}
d\left(\theta\right) &= d_1 \cos^2 \left(\theta - \theta_0 \right) + d_2 \sin^2 \left(\theta - \theta_0 \right) \\
&=\frac{d_1+d_2}{2} + \frac{d_1-d_2}{2} \cos \left( 2 \left(\theta - \theta_0 \right)\right)
\end{align}

where $d_1, d_2$ and $\theta_0$ allow the complete tensor to be reconstructed. This corresponds to the cross-section of an ellipsoid in general. The formula can be rigorously justified for the Maxwell equations by considering the first-order electric susceptibility as a rank-2 polar tensor in the relevant monoclinic symmetry, and considering the imaginary part as a perturbative absorption behavior with respect to refractive aspects driven by the real part. \cite{petit2013recent}.
In \eqref{eq:dinndorf}, if one finds $\theta_0=0$, then the principal axes of the investigated absorption transition are aligned with those of the dielectric frame, showing diagonal values of $d_1$ and $d_2$. As shown in Fig.\ref{fig:ErYSO_all_article}, this is clearly not the case, highlighting here the monoclinic behavior of \yso. The four data set in Fig.\ref{fig:ErYSO_all_article} are fitted by \eqref{eq:dinndorf} (dashed-dotted curves) to obtain the results reported \diff{the table \ref{tab:fit}}.

\begin{table}[htbp]
\centering
\caption{ Fitting parameters}
\begin{tabular}{lccc}
\toprule
Experimental  set & $d_1$ & $d_2$ & $\theta_0$ \\
\midrule
Site 1 - polar. along  {\bf b} (blue) & $1.60$ & $0.55$ & $115^\circ$ \\
Site 1 - polar. in  ({\bf D$_1$,D$_2$}) plane (red)& $1.59$ & $1.20$ & $18^\circ$ \\
Site 2 - polar. along  {\bf b} (blue) & $0.98$ & $0.09$ & $68^\circ$ \\
Site 2 - polar. in  ({\bf D$_1$,D$_2$}) plane (red) & $1.29$ & $1.27$ & $96^\circ$ \\
\bottomrule
\end{tabular}
  \label{tab:fit}
\end{table}

The two ideal cases independently with either a purely ED or MD transition are considered below.

(i) If the transition was purely ED (so called forced-ED), our configuration would be sufficient to extract the full transition tensor for both sites. Because {\bf b} is an axis of symmetry, when the electric field polarization stands the ordinary polarization mode along {\bf b}, the absorption should be constant to a value $d_b^E$ as $\theta$ is varied: the blue curves in Fig.\ref{fig:ErYSO_all_article} would be flat. In the ({\bf D$_1$,D$_2$}) plane, there is no reason for {\bf D$_1$} nor {\bf D$_2$} to be the absorption eigenaxes, so that the absorption angular distribution should be characterized by three values $d_1^E$, $d_2^E$ and $\theta_0^E$. The latter is to be non-zero due to the existence of off-diagonal components in such dielectric plane orthogonal to the monoclinic {\bf b}-axis.
As absorption is not constant while considering a fixed polarization mode along {\bf b}, it strongly evidences that the transition is not purely ED. This is expected because the $^{4}I_{15/2} \rightarrow $$^{4}I_{13/2}$ transition is MD allowed (with $\Delta J = -1$ as selection rule \cite{li1992spectroscopic}).

(ii) If the transition was purely MD, we can follow the previous reasoning for ED transition replacing the electric field by the magnetic field polarization, these fields being typically perpendicular to the propagation axis $\vec{k}$ and to each other. Namely, if the electric field lies in the  ({\bf D$_1$,D$_2$}) plane, then the magnetic field is along {\bf b} and the MD absorption should be constant $d_b^H$:  the red curves in Fig.\ref{fig:ErYSO_all_article} would be flat. By analogy,  if the electric field is along  {\bf b}, then the absorption angular distribution should be characterized by the three values $d_1^H$, $d_2^H$ and $\theta_0^H$. \diff{From the fact that the red curve for site 2 in Fig.\ref{fig:ErYSO_all_article} is flat, it is extremely tempting to conclude that the site has a dominant MD character. One cannot nevertheless discard the fact that the ED components are accidentally isotropic in the ({\bf D$_1$,D$_2$}) plane, leading to an apparent purely MD transition pattern. Further analysis is needed to lift the ambiguity.}

If now is considered a mixture of ED and MD transitions, then the ED and MD absorptions have to be simply added. Actually, the  measurements do stand in the weak absorption approximation and we measure the linear absorption where ED and MD can be taken independently. Following the previous reasoning, the angular distribution when the electric field polarization is along {\bf b} can be written as
\begin{equation}\label{eq:along_X}
d_\mathrm{blue}\left(\theta\right) = \underbrace{d_b^E + \frac{d_1^H+d_2^H}{2}}_{\frac{d_1+d_2}{2}} + \underbrace{\frac{d_1^H-d_2^H}{2}}_{\frac{d_1-d_2}{2}} \cos ( 2 (\theta -\underbrace{ \theta_0^H}_{\theta_0} ))
\end{equation}
And when the electric field lies in the ({\bf D$_1$,D$_2$}) plane, it comes:
\begin{equation}\label{eq:in_YZ}
d_\mathrm{red}\left(\theta\right) = \underbrace{d_b^H + \frac{d_1^E+d_2^E}{2}}_{\frac{d_1+d_2}{2}} + \underbrace{\frac{d_1^E-d_2^E}{2}}_{\frac{d_1-d_2}{2}} \cos ( 2 (\theta - \underbrace{\theta_0^E}_{\theta_0} ))
\end{equation}
In order to determine unambiguously the components of the ED and MD absorption tensors,it is necessary to have 4 values per curve, i.e. $d_b^E,d_1^H,d_2^H,\theta_0^H$ for the blue and $d_b^H,d_1^E,d_2^E,\theta_0^E$ for the red one, while the fit gives only 3 parameters: $d_1$,$d_2$ and $\theta_0$ reported as underbraces in \eqref{eq:along_X} and \eqref{eq:in_YZ}. So we can only take the values in table \ref{tab:fit} as phenomenological parameters to describe the absorption in the  ({\bf D$_1$,D$_2$}) measurement plane (with two polarizations) but without allowing for a complete determination of the absorption tensors. This is a consequence of the ED and MD mixture of the  $^{4}I_{15/2} \rightarrow $$^{4}I_{13/2}$  transition of \er in \yso. A complete determination would require to repeat the measurement in different planes as ({\bf b,D$_1$}) and ({\bf b,D$_2$}), thus giving redundantly the ED and MD tensors values.

\section{Conclusion}
We perform for the first time to the best of our knowledge a low temperature measurement of the absorption angular distribution in polarized light in the plane perpendicular to the monoclinic special axis of \yso, which allows us to spectrally resolve and independently characterize the two yttrium substitution sites. Measurements remarkably demonstrate that this transition of interest, namely the  $^{4}I_{15/2} \rightarrow $$^{4}I_{13/2}$  transition between the lowest crystal field levels of \er, actively investigated for quantum information processing, is not only a magnetic-dipole allowed transition but indeed a hybrid electric-magnetic transition. Because of such behavior of ED and MD mixture, the absorption tensor components cannot fully be unambiguously determined yet while only considering the currently investigated dielectric plane. The use of a sphere instead of a cylinder would ideally permit to explore the different independent measurement planes by following the procedure from \cite{petit2013recent}. But it would be very difficult to rotate a sphere inside a cryostat. The use of three oriented cylinders should then be used, but at the price of tedious crystal machining and polishing.

Nevertheless, one should note that recent progress in determining the crystal field parameters in the low symmetry site of \yso  \cite{PhysRevLett.123.057401} make possible a calculation of the allowed MD component, as opposed to the ED component which is forbidden or so-called forced. Complemented by a theoretical prediction of the MD component, our measurements  would be sufficient to fully characterize the forced ED and allowed MD tensors.

To conclude with the spin Hamiltonian analogy outlined in the introduction, it would be interesting to repeat absorption measurements under magnetic field. One would obtain the equivalent of the g-tensor that can be extracted from electron spin resonance spectroscopy. It may sounds surprising that the g-tensor has 6 independent components \cite{sun_magnetic_2008} while the ED and MD optical transition tensors only have 4 independent components each here. This is precisely because we do not apply a magnetic field that the monoclinic symmetry (and not the local $C_1$ symmetry) strictly applies in the sense that {\bf b} stays an axis of symmetry. Under magnetic field, the optical transition tensors should have 6 independent components because {\bf b} may not remain an axis of symmetry, forcing the magnetically-stressed crystal to behave similarly to a triclinic crystal.
One could alternatively reinterpret this global symmetry analysis as lifting the inequivalence between the so-called magnetic sub-sites, related by the crystal symmetry, whose degeneracy can be precisely lifted by applying a magnetic field, away from the {\bf b} axis or the ({\bf D$_1$,D$_2$}) plane.
Such experiment is quite challenging because the crystal has to be rotated with the magnetic field in the cryostat. The latter should be sufficiently strong to lift the degeneracy and resolve the Zeeman transition. This is an interesting perspective.

\section{Funding Information}

We have received funding from the Investissements d'Avenir du LabEx PALM ExciMol, ATERSIIQ and OptoRF-Er (ANR-10-LABX-0039-PALM). This work was supported by the ANR MIRESPIN project, grant ANR-19-CE47-0011 of the French Agence Nationale de la Recherche.

\section{Acknowledgments}

We thank A. Kinos for enlightening discussions and P. Goldner, M. Reid and C. Thiel for their stimulating interest in our work.

\bibliographystyle{cas-model2-names}
\bibliography{abs_cylindre_ErYSO_bib}

%


\end{document}